# Synaptic sampling: A connection between PSP variability and uncertainty explains neurophysiological observations

Laurence Aitchison and Peter E. Latham

July 12, 2015

## Abstract

When an action potential is transmitted to a postsynaptic neuron, a small change in the postsynaptic neuron's membrane potential occurs. These small changes, known as a postsynaptic potentials (PSPs), are highly variable, and current models assume that this variability is corrupting noise. In contrast, we show that this variability could have an important computational role: representing a synapse's uncertainty about the optimal synaptic weight (i.e. the best possible setting for the synaptic weight). We show that this link between uncertainty and variability, that we call synaptic sampling, leads to more accurate estimates of the uncertainty in task relevant quantities, leading to more effective decision making. Synaptic sampling makes three predictions, all of which have some experimental support. First the more variable a synapse is, the more it should change during LTP protocols. Second, variability should increase as the presynpatic firing rate falls. Third, PSP variance should be proportional to PSP mean. We provide support for the first two predictions by reanalysing existing datasets, and we find preexisting data in support of the last two predictions.

## 1 Introduction

To transmit an action potential to a downstream neuron, the presynaptic terminal releases one or more vesicles of neurotransmitter [1]. The neurotransmitter diffuses across the synaptic cleft and opens postsynaptic ion chanels, causing a change in the postsynaptic membrane potential, known as a PSP [1]. However, the number of vesicles released by a single presynaptic terminal is highly variable, and can even be zero [2], giving rise to a similarly high level of variability in PSP amplitudes.

Previously, this variability has been ignored, treated as corrupting noise, or regarded as a mechanism to mildly, and only in limited circumstances, increase the amount of information transmitted per vesicle release event [3, 4, 5, 6]. In contrast, we show that variability could have an important computational role: representing uncertainty [7, 8]. In particular, we give a new hypothesis, synaptic sampling, which states, loosely, that more variable PSPs represent more uncertainty.

Uncertainty is involved in almost every judgement and decision we take. For instance, it is impossible to know exactly how long a journey will take, especially when using public transportation. It is only possible to give an estimate, which may state the degree of uncertainty, for instance, "I should be 15 minutes, but it could take an hour". In the laboratory, humans have been shown to use information about uncertainty when solving a variety of sensory [9, 10, 11, 12], motor [13, 14] and cognitive [15, 16, 17] tasks.

Importantly, the brain is uncertain not only about predictions and the current state of the world, but also about the optimal synaptic weights (or, equivalently, the optimal PSPs). The optimal synaptic weights are simply the set of weights that ensure that the circuit, system, and ultimately organism functions as effectively as possible. The brain cannot be certain about these optimal synaptic weights because there are many, many synapses (around $10^{15}$) and relatively little data. In particular, even using very generous estimates of the amount of information entering the brain ($10^6$ bits per second), it would take 30 years to collect one bit of information about each optimal weight — which would still leave considerable uncertainty.

However, it is not straightforward for synapses to estimate their uncertainty. In the companion to this paper [18] we derive neurally plausible Bayesian learning rules which estimate uncertainty in synaptic weights, and use those estimates to learn more rapidly than is possible using classical learning rules.

Importantly, estimates of uncertainty found by the Bayesian learning rules can be used to do more than just speed up synaptic learning. The critical insight is that task relevant quantities are computed by combining sensory input with synaptic weights. Uncertainty about synaptic weights therefore induces uncertainty in task relevant quantities. However, under Bayesian learning rules alone, estimates of uncertainty in synaptic weights are simply used to set the learning rate. Estimates of uncertainty therefore cannot affect neural activity on short timescales, and so cannot contriubte to estimates of the uncertainty in task relevant quantities. Here, we propose a mechanism, synaptic sampling, by which uncertainty in a synaptic weight could affect downstream neural activity, and therefore allow the circuit to estimate uncertainty in task relevant quantities. In particular, synaptic sampling states that as a synapse becomes more uncertain about the synaptic weight, PSP amplitudes become more variable. This variability propagates to downstream neural activity, and therefore can be used to represent uncertainty about task relevant quantities in downstream circuits. Thus, under our model, variability in PSP amplitudes is a feature that allows accurate estimation of uncertainty, not a bug.

This idea, that variability in neural activity (which in our case is induced by synaptic sampling), can be used to represent uncertainty is known as the sampling hypothesis [7, 8, 19]. Concretely, the sampling hypothesis states that neural activity represents the state of the world, there is uncertainty about the state of the world, and the degree of uncertainty is represented by the degree of variability in neural activity. The sampling hypothesis is supported by some experimental data [19], and in general, sampling has considerable computational advantages over other neural representations of uncertainty [20].

The Results section is divided into two parts. It begins by describing the advantages of synaptic sampling, and then moves on to the three predictions made by synaptic sampling, all of which have some experimental support. First, more variable synapses change more during LTP protocols. Second, we have not yet tested the prediction that variability should decrease as average presynaptic firing rate increases. Third, PSP mean is proportional to PSP variance.

## 2 Results

Synaptic sampling (and, analogously, the sampling hypothesis) actually make a stronger statement than "variability increases with uncertainty". Synaptic sampling states that synapses both maintain a distribution over the optimal synaptic weight, and when there is a presynaptic spike, the PSP is drawn from that distribution. Therefore, the PSP mean and variance are just the mean and variance of the distribution over the optimal synaptic weights, {eq:def:ss}

$$\text{PSP mean} = \mathrm{E}\left[w_{\text{opt},i}|\text{Training Data}\right], \tag{1a}$$
$$\text{PSP variance} = \mathrm{Var}\left[w_{\text{opt},i}|\text{Training Data}\right], \tag{1b}$$

where, $w_{\text{opt},i}$ is the optimal weight, the weight that ensures the circuit functions as effectively as possible, and $i$ denotes the $i$th presynaptic cell. So, not only does variability increase with uncertainty, but there is a very specific relationship between variability and uncertainty. As we will show later, having this relationship is necessary to get the computational advantages of synaptic sampling.

### 2.1 Inference and action selection with uncertain weights

To select the correct action, knowing the uncertainty in task relevant quantities is critical. For instance, to decide whether to jump over a puddle, it is important to have not only a central estimate of landing location, relative to the end of the puddle), but also the uncertainty, or standard deviation in the estimate. Uncertainty about the landing location comes from two sources, uncertainty about the current state of the world and uncertainty about the optimal weights. To see how the brain might compute uncertainty in landing location, we consider a simplified scenario, in which we use $\mathbf{x}_{\text{opt}}$ to denote the best possible distributed, spike-based representation of the true state of the external world. Of course, $\mathbf{x}_{\text{opt}}$ represents the true state of the external world, and as the brain does not know the true state of the external world it cannot know $\mathbf{x}_{\text{opt}}$. As $\mathbf{x}_{\text{opt}}$ is a spike-based representation, it is a binary vector where $x_{\text{opt},i} = 1$ indicates a "spike". The optimal estimate of the landing location, denoted $y_{\text{opt}}$, is then given by

$$y_{\text{opt}} = \mathbf{w}_{\text{opt}} \cdot \mathbf{x}_{\text{opt}} + \gamma_y \eta. \tag{2}$$

where $\eta$ is standard Gaussian noise, $P(\eta) = \mathcal{N}(0,1)$, representing the small amount of uncertainty about landing location that remains when $\mathbf{x}_{\text{opt}}$ and $\mathbf{w}_{\text{opt}}$ are known precisely.

Of course, the brain knows neither the optimal weights, $\mathbf{w}_\text{opt}$, nor, as mentioned, the true state of the external world, $\mathbf{x}_\text{opt}$. Instead, the brain could compute a (possibly noisy) "best guess" of $\mathbf{x}_\text{opt}$, and the neuron could use a "best guess" of $\mathbf{w}_\text{opt}$, resulting in

$$y_\text{best guess} = \mathbf{w}_\text{best guess} \cdot \mathbf{x}_\text{best guess} + \gamma_y \eta \qquad (3)$$

However, not only might these guesses give a poor mean estimate of landing location, this scheme is completely unable to give an estimate of uncertainty — so offers little guidance as to whether or not you should jump over the puddle.

In contrast, the estimate of landing location should be based only observations, not guesses. The synapses observes both current sensory data, which gives information about $\mathbf{x}_\text{opt}$, and past training data, which gives information about $\mathbf{w}_\text{opt}$. Ideally, the system would therefore like to know the distribution over landing locations conditioned on the sensory and training data,

$$P(y_\text{opt}|\text{Sensory Data}, \text{Training Data}), \qquad (4)$$

but it is difficult for neural circuits to compute this distribution directly. However, it is straightforward for neural circuits to draw samples, $y$, from this distribution,

$$y \sim P(y_\text{opt}|\text{Sensory Data}, \text{Training Data}), \qquad (5)$$

using synaptic sampling. To draw these samples, we simply need to set neural activity, $\mathbf{x}$, to a pattern that represents a plausible state of the world,

$$\mathbf{x} \sim P(\mathbf{x}_\text{opt}|\text{Sensory Data}), \qquad (6)$$

and set the synaptic weights, $\mathbf{w}$, to values that represent a plausible setting for the value of the optimal weights,

$$\mathbf{w} \sim P(\mathbf{w}_\text{opt}|\text{Training Data}). \qquad (7)$$

A sample of landing location is given by combining the sampled inputs and the sampled weights, which could be done by a single neuron,

$$y = \mathbf{w} \cdot \mathbf{x} + \gamma_y \eta. \qquad (8)$$

Remarkably, a single neuron can therefore draw samples from a distribution conditioned only on data that is actually known, without making assumptions (which would almost certainly be wrong) about either the state of the world or the optimal weights.

Our argument appears to assume that the brain uses the output of a single neuron to make predictions. This is not too implausible — the cerebellum does contain a large number of Purkinje cells [21] that are believed to use supervised learning to, amoung other things, make predictions (though perhaps not about landing location). However, it is certainly possible that such a computation is performed by a larger circuit. As long as that network is effectively feedforward, we can still, by the logic described above, estimate its uncertainty by combining synaptic sampling with the sampling hypothesis.

## 2.2 Predictions

Now, we discuss the predictions made by the synaptic sampling hypothesis.

### 2.2.1 Plasticity experiments

The learning rules [18] state that, under very general conditions, the percentage change in weight induced by spiking activity is related to the synapse's normalized uncertainty,

$$\text{Percentage change in weight} \propto \frac{\text{PSP Variance}}{\text{PSP Mean}} \equiv \text{Normalized Variability}. \qquad (9) \quad \{\texttt{eq:ltp-variability}\}$$

Intuitively, such an update rule makes sense: a synapse that is very certain about the value of the optimal weight should update its mean estimate only a little upon receiving new information. In contrast, a synapse that is very uncertain should make large updates upon receiving new information.

We begin using data taken from a single LTP protocol (to reduce variability associated with different protocols) and with strong LTP (50 Hz). We saw that, barring one outlier, there is a strong relationship between normalized variability and the change in the weight (Figure 1A). We would clearly like to assess the significance of this relationship. However, we cannot simply use linear regression, because our data violates two assumptions made in the usual process for obtaining p-values for linear regression. Firstly, linear regression assumes that the noise is Gaussian distributed, and thus that there are no outliers — of course, we do have an outlier. Second, linear regression assumes that the noise (i.e. variability in the percentage change) is the same, in contrast, in our data, it appears that the noise increases as the value on the x-axis (normalized variability) increases. This increase in noise is expected, because we expect the size of the feedback signal, $f$, and the normalizer, $s_{\text{like}}^2$, (and perhaps other factors) to vary between cells, changing the slope of the relationship between percentage change and normalised variability. Thus, we developed a model, described in the Supplementary Material, which takes into account not only outliers and increasing variance, but also any dependence of the percentage change on the mean, and gives ($p < 0.001$, one-sided).

We expect that combining data from a wider range of protocols will give a similar pattern, but with higher variability in slopes. Indeed, taking data from a wider range of high-frequency stimulation protocols (40 Hz, 50 Hz or 100 Hz) gives a similar pattern (outliers and increasing variance), though with a far wider range of slopes (Figure 1B). The addditional variability in this data reduces the significance, ($p < 0.05$)

### 2.2.2 Presynaptic firing rates

A mean-field analysis [18], tells us that the variability should depend on the presynaptic firing rate, and, in particular, the normalized variability should fall

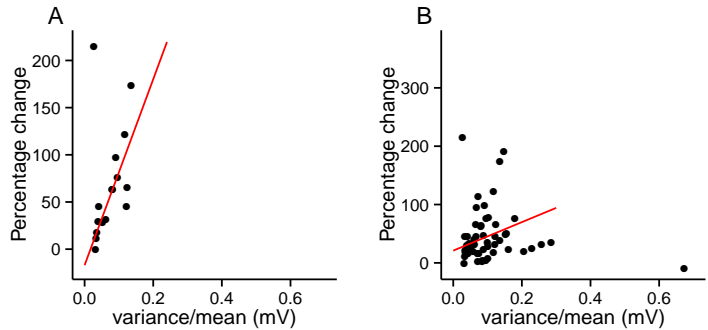

Figure 1: The percentage change in weight during an LTP protocol increases as the noise level increases. The red line is our fit. **A** All data with frequency 30 Hz or above, corresponding to strong LTP. **B** Data from one common protocol, at 50 Hz. Data from [22, 23]. {fig:ltp}

as the presynaptic firing rate rises,

$$\frac{\text{Normalized}}{\text{Variability}} = \frac{\text{PSP Variance}}{\text{PSP Mean}} \propto \frac{1}{\sqrt{\text{Presynaptic firing rate}}}. \qquad (10) \quad \{\texttt{eq:rate}\}$$

Intuitively, higher presynaptic firing rates give the synapse more opportunities to update the synaptic weight, allowing the synapse to become more certain, and hence reducing variability.

We therefore took data from [24], in which they recorded Calcium signals in V1 *in vivo* under a variety of stimulation conditions, giving an estimate of firing rate, then were able to patch the same cells *in vitro*, in order to get the mean and variance of PSPs. Again, the gradient of the trend line is significantly different from 0 ($p < 0.003$, regressing log variance against log mean and log rate jointly, and reporting the p-value for the rate, to ensure that any dependence of the mean on the firing rate does not contaminate our results), but not significantly different from our model prediction.

### 2.2.3 Variance is proportional to the mean

Equation (10) suggests that, if we average over firing rates, PSP variance should be proportional to PSP mean. It is possible to test this prediction using much larger datasets than the one used above, because we do not now require corresponding *in vivo* calcium imaging data. In particular, we used data from [26].

## 3 Discussion

We showed that synaptic sampling allows neural systems to take uncertainty about synaptic weights into account during decision making — producing more accurate estimates of uncertainty and hence better decisions. Furthermore, synaptic sampling makes three predictions, three of which are consistent with experimental data, and one that remains to be tested. First, the percentage change in weight during an LTP experiment increases with variability — notably, we needed to reanalyse existing data to confirm this prediction. Second, variability increases as the presynaptic firing rate falls. Third, PSP variance is proportional to PSP mean.

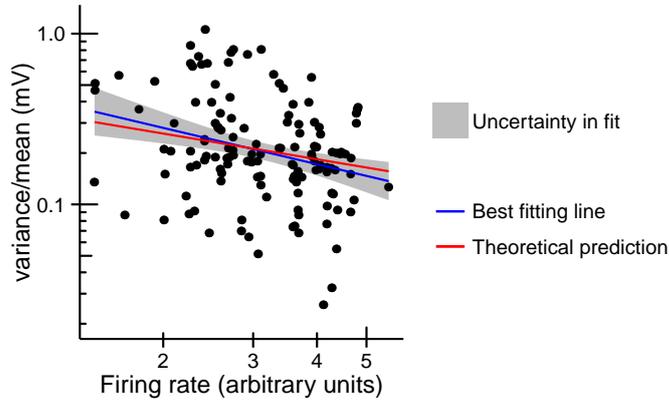

Figure 2: Normalized variability falls as firing rate increases. The blue line is fit by linear regression, and the grey region represents 2 standard errors. The red line, which has a slope of $-1/2$, is our prediction. Firing rate was measured by taking the average signal from a spike deconvolution algorithm [25]. While this is a highly approximate, noisy measure, it should bear some relationship to the true average presynaptic firing rate. It is also possible to use alternative measures of the firing rate, like the number of times the signal passes above a threshold, which still give significant results. Data from [24]. {fig:rate}

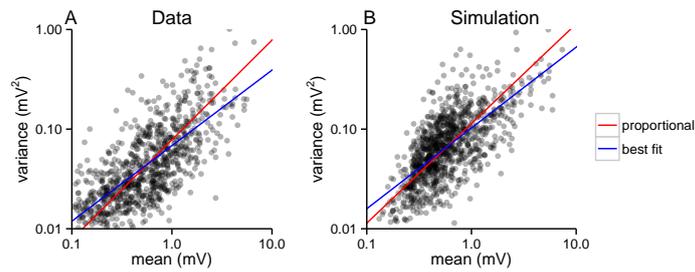

Figure 3: The variance is roughly proportional to the mean, both in data **A** and simulation **B**. The blue line was fitted by linear regression, and the red line has a slope of 1, representing proportionality. Data from [26]. {fig:mv}

Synaptic sampling states not only that normalized variability increases with normalized uncertainty, but also specifies the exact relationship between those quantities. However, our data is consistent with another closely related hypothesis. Our hypothesis, synaptic sampling, states that normalized variability equals normalized uncertainty (or equivalently that the variance of the distribution over the optimal weights is equal to the variance of the PSPs). An alternative hypothesis would be that normalized variability is proportional to normalized uncertainty (or equivalently, that the variances are proportional). Functionally, such a relationship could arise because reducing PSP variability costs energy, in which case, it might be worth expending additional energy to get low variability only when the optimal weight is known precisely. In contrast, when there is considerable uncertainty about the synaptic weight, it may not be as problematic to use highly variable PSPs. To distinguish between these competing hypotheses, we would need to determine the constant of proportionality. This is not currently possible, as it requires an extremely detailed characterisation of a neuron and its presynaptic partners. However, such an analysis should become possible within the next decade.

Our work has considerable implications in three areas. First, the experimental results presented in this paper provide strong, though by no means conclusive, support for the Bayesian learning rules presented in [18] — suggesting that the brain is Bayesian at the lowest level. Second, if the brain does indeed use synaptic sampling, this suggests that the brain's central mechanism for representing and computing with uncertainty is sampling. This is because, as we saw earlier, combining uncertainty (e.g. about the state of the world and the synaptic weights) by combining samples is trivial, while there is currently no mechanism that can be used to combine sampling with, for instance, a probabilistic population code [27, 28, 29]. Finally, it appears that variability gives us a way to probe the synapse's degree of uncertainty about its synaptic weight, which, we hope, will inspire novel experimental techniques.

# Acknowledgements

We would like to acknowledge Jesper Sjöstrum for data, and for valuable feedback on this manuscript. This work was supported by the Gatsby Charitable Foundation.

# Methods

We developed a customised statistical test for the data in Figure 1. We were required to do this because standard parametric statistical tests are based on a linear Gaussian model, and our data violates two assumptions made by these methods. First, the data should contain no outliers — our data clearly contains outliers. Second, the variance should be constant as the value on the x-axis increases — again this is clearly not true for our data.

We therefore used these features to write down a sensible model for our data. This model states that the data point is an outlier with probability 0.01,

$$P(o_i) = \text{Bernoulli}\left(o_i; 0.01\right) \tag{11}$$

If the data point is an outlier, then it is drawn from a broad Gaussian,

$$P\left(\frac{\Delta m_i}{m_i}|o_i = 1\right) = \mathcal{N}\left(\frac{\Delta m_i}{m_i}; 0, 2\right). \tag{12}$$

In contrast, if the data point is not an outlier, then it is drawn from Gaussian, whose mean and variance both depend on a scaled version of the normalized variability,

$$r_i = \frac{\frac{s_i^2}{m_i}}{\left\langle \frac{s_i^2}{m_i} \right\rangle}. \tag{13}$$

In particular,

$$P\left(\frac{\Delta m_i}{m_i}|o_i = 0\right) = \mathcal{N}\left(\frac{\Delta m_i}{m_i}; ar_i + b + cm_i, (d + er_i)^2\right) \tag{14}$$

Note the inclusion of $c$, which accounts for any dependence of the percentage change on the mean weight. Furthermore, note that we have represented the variance as a sum of two contributions, one constant, and one dependent on $r_i$. In order to identify $d$ and $e$, which controlled the variance, we had to specify a prior (though elsewhere, for $a$, $b$ and $c$, we used uniform, non-informative priors). We used the most standard prior for a parameter controlling the variance, a Gamma distribution, and we used mean 0.2,

$$P(d) = \text{Gamma}\left(d; 2, 10\right), \tag{15}$$
$$P(e) = \text{Gamma}\left(d; 2, 10\right). \tag{16}$$

We inferred the parameters, using the widespread, industry-standard MCMC program, STAN. Our significance levels come from the confidence intervals returned by STAN.